\documentclass[12pt,a4paper]{article}
\usepackage{epsfig}
\usepackage{amsmath}
\usepackage{hhline}
\usepackage{amssymb}
\usepackage{times}
\usepackage{cite}
\usepackage{multirow}
\usepackage{lscape}
\usepackage{xtab}
\usepackage{url}

\newlength{\dinwidth}
\newlength{\dinmargin}
\begin{document}  
\newcommand{\pom}{{I\!\!P}}
\newcommand{\reg}{{I\!\!R}}
\newcommand{\slowpi}{\pi_{\mathit{slow}}}
\newcommand{\fiidiii}{F_2^{D(3)}}
\newcommand{\fiidiiiarg}{\fiidiii\,(\beta,\,Q^2,\,x)}
\newcommand{\n}{1.19\pm 0.06 (stat.) \pm0.07 (syst.)}
\newcommand{\nz}{1.30\pm 0.08 (stat.)^{+0.08}_{-0.14} (syst.)}
\newcommand{\fiidiiiful}{F_2^{D(4)}\,(\beta,\,Q^2,\,x,\,t)}
\newcommand{\fiipom}{\tilde F_2^D}
\newcommand{\ALPHA}{1.10\pm0.03 (stat.) \pm0.04 (syst.)}
\newcommand{\ALPHAZ}{1.15\pm0.04 (stat.)^{+0.04}_{-0.07} (syst.)}
\newcommand{\fiipomarg}{\fiipom\,(\beta,\,Q^2)}
\newcommand{\pomflux}{f_{\pom / p}}
\newcommand{\nxpom}{1.19\pm 0.06 (stat.) \pm0.07 (syst.)}
\newcommand {\gapprox}
   {\raisebox{-0.7ex}{$\stackrel {\textstyle>}{\sim}$}}
\newcommand {\lapprox}
   {\raisebox{-0.7ex}{$\stackrel {\textstyle<}{\sim}$}}
\def\gsim{\,\lower.25ex\hbox{$\scriptstyle\sim$}\kern-1.30ex%
\raise 0.55ex\hbox{$\scriptstyle >$}\,}
\def\lsim{\,\lower.25ex\hbox{$\scriptstyle\sim$}\kern-1.30ex%
\raise 0.55ex\hbox{$\scriptstyle <$}\,}
\newcommand{\pomfluxarg}{f_{\pom / p}\,(x_\pom)}
\newcommand{\dsf}{\mbox{$F_2^{D(3)}$}}
\newcommand{\dsfva}{\mbox{$F_2^{D(3)}(\beta,Q^2,x_{I\!\!P})$}}
\newcommand{\dsfvb}{\mbox{$F_2^{D(3)}(\beta,Q^2,x)$}}
\newcommand{\dsfpom}{$F_2^{I\!\!P}$}
\newcommand{\gap}{\stackrel{>}{\sim}}
\newcommand{\lap}{\stackrel{<}{\sim}}
\newcommand{\fem}{$F_2^{em}$}
\newcommand{\tsnmp}{$\tilde{\sigma}_{NC}(e^{\mp})$}
\newcommand{\tsnm}{$\tilde{\sigma}_{NC}(e^-)$}
\newcommand{\tsnp}{$\tilde{\sigma}_{NC}(e^+)$}
\newcommand{\st}{$\star$}
\newcommand{\sst}{$\star \star$}
\newcommand{\ssst}{$\star \star \star$}
\newcommand{\sssst}{$\star \star \star \star$}
\newcommand{\tw}{\theta_W}
\newcommand{\sw}{\sin{\theta_W}}
\newcommand{\cw}{\cos{\theta_W}}
\newcommand{\sww}{\sin^2{\theta_W}}
\newcommand{\cww}{\cos^2{\theta_W}}
\newcommand{\trm}{m_{\perp}}
\newcommand{\trp}{p_{\perp}}
\newcommand{\trmm}{m_{\perp}^2}
\newcommand{\trpp}{p_{\perp}^2}
\newcommand{\alp}{\alpha_s}

\newcommand{\alps}{\alpha_s}
\newcommand{\sqrts}{$\sqrt{s}$}
\newcommand{\LO}{$O(\alpha_s^0)$}
\newcommand{\Oa}{$O(\alpha_s)$}
\newcommand{\Oaa}{$O(\alpha_s^2)$}
\newcommand{\PT}{p_{\perp}}
\newcommand{\JPSI}{J/\psi}
\newcommand{\sh}{\hat{s}}
\newcommand{\uh}{\hat{u}}
\newcommand{\MP}{m_{J/\psi}}
\newcommand{\PO}{I\!\!P}
\newcommand{\xbj}{x}
\newcommand{\xpom}{x_{\PO}}
\newcommand{\ttbs}{\char'134}
\newcommand{\xpomlo}{3\times10^{-4}}  
\newcommand{\xpomup}{0.05}  
\newcommand{\dgr}{^\circ}
\newcommand{\pbarnt}{\,\mbox{{\rm pb$^{-1}$}}}
\newcommand{\gev}{\,\mbox{GeV}}
\newcommand{\WBoson}{\mbox{$W$}}
\newcommand{\fbarn}{\,\mbox{{\rm fb}}}
\newcommand{\fbarnt}{\,\mbox{{\rm fb$^{-1}$}}}
\newcommand{\dsdx}[1]{$d\sigma\!/\!d #1\,$}
\newcommand{\eV}{\mbox{e\hspace{-0.08em}V}}
%
%
\newcommand{\qsq}{\ensuremath{Q^2} }
\newcommand{\gevsq}{\ensuremath{\mathrm{GeV}^2} }
\newcommand{\et}{\ensuremath{E_t^*} }
\newcommand{\rap}{\ensuremath{\eta^*} }
\newcommand{\gp}{\ensuremath{\gamma^*}p }
\newcommand{\dsiget}{\ensuremath{{\rm d}\sigma_{ep}/{\rm d}E_t^*} }
\newcommand{\dsigrap}{\ensuremath{{\rm d}\sigma_{ep}/{\rm d}\eta^*} }

\newcommand{\dstar}{\ensuremath{D^*}}
\newcommand{\dstarp}{\ensuremath{D^{*+}}}
\newcommand{\dstarm}{\ensuremath{D^{*-}}}
\newcommand{\dstarpm}{\ensuremath{D^{*\pm}}}
\newcommand{\zDs}{\ensuremath{z(\dstar )}}
\newcommand{\Wgp}{\ensuremath{W_{\gamma p}}}
\newcommand{\ptds}{\ensuremath{p_t(\dstar )}}
\newcommand{\etads}{\ensuremath{\eta(\dstar )}}
\newcommand{\ptj}{\ensuremath{p_t(\mbox{jet})}}
\newcommand{\ptjn}[1]{\ensuremath{p_t(\mbox{jet$_{#1}$})}}
\newcommand{\etaj}{\ensuremath{\eta(\mbox{jet})}}
\newcommand{\detadsj}{\ensuremath{\eta(\dstar )\, \mbox{-}\, \etaj}}

\newcommand{\xsw}{\ensuremath{\sigma_{W}}}
\newcommand{\xsisolep}{\ensuremath{\sigma_{\ell+{P}_{T}^{\rm miss}}}}

\newcommand{\dkl}{\ensuremath{\left(\Delta\kappa,\lambda\right)}}
\newcommand{\dkldk}{\ensuremath{\left(\Delta\kappa,\lambda=0\right)}}
\newcommand{\dkll}{\ensuremath{\left(\Delta\kappa=0,\lambda\right)}}
\newcommand{\dklp}{\ensuremath{\left(\Delta\kappa^{\prime},\lambda^{\prime}\right)}}

\newcommand{\cosths}{\ensuremath{\cos\,\theta^{*}}}
\newcommand{\qcosths}{\ensuremath{q_{\ell}\cdot\cosths}}

\def\Journal#1#2#3#4{{#1} {\bf #2} (#3) #4}
\def\NCA{\em Nuovo Cimento}
\def\NIM{\em Nucl. Instrum. Methods}
\def\NIMA{{\em Nucl. Instrum. Methods} {\bf A}}
\def\NPB{{\em Nucl. Phys.}   {\bf B}}
\def\PLB{{\em Phys. Lett.}   {\bf B}}
\def\PRL{\em Phys. Rev. Lett.}
\def\PRD{{\em Phys. Rev.}    {\bf D}}
\def\ZPC{{\em Z. Phys.}      {\bf C}}
\def\EJC{{\em Eur. Phys. J.} {\bf C}}
\def\CPC{\em Comp. Phys. Commun.}

\begin{titlepage}

\noindent
\begin{flushleft}
{\tt DESY 09-140    \hfill    ISSN 0418-9833} \\
{\tt September 2009}\\ 
\end{flushleft}

\vspace{2cm}
\begin{center}
\begin{Large}

{\bf Events with an Isolated Lepton and Missing Transverse Momentum and
Measurement of {\boldmath $W$} Production at HERA}

\vspace{2cm}

The H1 and ZEUS Collaborations

\end{Large}
\end{center}

\vspace{2cm}

\begin{abstract}
\noindent

A search for events containing an isolated electron or muon and
missing transverse momentum produced in $e^{\pm}p$ collisions is
performed with the H1 and ZEUS detectors at HERA.
The data were taken in the period $1994$--$2007$ and correspond to an
integrated luminosity of $0.98$~fb$^{-1}$.
The observed event yields are in good overall agreement with the
Standard Model prediction, which is dominated by single $W$
production.
In the $e^{+}p$ data, at large hadronic transverse momentum
$P_{T}^{X}>25$~GeV, a total of $23$ events are observed compared to a
prediction of $14.0 \pm 1.9$.
The total single $W$ boson production cross section is measured as
$1.06 \pm 0.16~({\rm stat.}) \pm 0.07~({\rm sys.})$~pb, in agreement
with an SM expectation of $1.26 \pm 0.19$~pb.

\end{abstract}

\vspace{1.5cm}

\begin{center}
Accepted by JHEP
\end{center}

\end{titlepage}

\begin{flushleft}
  \begin{center}                                                                                     
{                      \Large  The H1 and ZEUS Collaborations              }                               
\end{center}                                                                                       
F.D.~Aaron$^{13, a8}$,
H.~Abramowicz$^{72, a36}$,
I.~Abt$^{57}$,
L.~Adamczyk$^{19}$,
M.~Adamus$^{84}$,
M.~Aldaya~Martin$^{31}$,
C.~Alexa$^{13}$,
K.~Alimujiang$^{31}$,
V.~Andreev$^{54}$,
S.~Antonelli$^{9}$
% ...

{\it
\vspace{0.4cm}

 $^{1}$  I. Physikalisches Institut der RWTH, Aachen, Germany \\
% ...
$^{9}$   {\it University and INFN Bologna, Bologna, Italy}~$^{b17}$ \\
% ...
$^{13}$  National Institute for Physics and Nuclear Engineering (NIPNE), Bucharest, Romania \\
% ...
$^{19}$   {\it Faculty of Physics and Applied Computer Science, AGH-University of Science and \mbox{Technology}, Cracow, Poland}~$^{b27}$ \\
% ...
$^{31}$   {\it Deutsches Elektronen-Synchrotron DESY, Hamburg, Germany} \\
% ...
$^{54}$  Lebedev Physical Institute, Moscow, Russia~$^{b6}$ \\
% ...
$^{57}$   {\it Max-Planck-Institut f\"ur Physik, M\"unchen, Germany} \\
% ...
$^{72}$   {\it Raymond and Beverly Sackler Faculty of Exact Sciences, School of Physics, Tel Aviv University, Tel Aviv, Israel}~$^{b16}$ \\
% ...
$^{84}$   {\it Institute for Nuclear Studies, Warsaw, Poland} \\
 $^{85}$  Fachbereich C, Universit\"at Wuppertal, Wuppertal, Germany \\
 $^{86}$  Yerevan Physics Institute, Yerevan, Armenia \\
 $^{87}$   {\it Meiji Gakuin University, Faculty of General Education, Yokohama, Japan}~$^{b18}$ \\
 $^{88}$   {\it Department of Physics, York University, Ontario, Canada M3J1P3}~$^{b14}$ \\
 $^{89}$   {\it Deutsches Elektronen-Synchrotron DESY, Zeuthen, Germany} \\
 $^{90}$  Institut f\"ur Teilchenphysik, ETH, Z\"urich, Switzerland~$^{b10}$ \\
 $^{91}$  Physik-Institut der Universit\"at Z\"urich, Z\"urich, Switzerland~$^{b10}$ \\

\vspace{0.4cm}
{ \small


$^{a1}$  Also at Physics Department, National Technical University, Zografou Campus, GR-15773 Athens, Greece \\
$^{a2}$  Also at Rechenzentrum, Universit\"at Wuppertal, Wuppertal, Germany \\
$^{a3}$  Also at University of P.J. \v{S}af\'arik, Ko\v{s}ice, Slovak Republic \\
$^{a4}$  Also at CERN, Geneva, Switzerland \\
$^{a5}$  Also at Max-Planck-Institut f\"ur Physik, M\"unchen, Germany \\
$^{a6}$  Also at Comenius University, Bratislava, Slovak Republic \\
$^{a7}$  Also at DESY and University Hamburg, Helmholtz Humboldt Research Award \\
$^{a8}$  Also at Faculty of Physics, University of Bucharest, Bucharest, Romania \\
% ...
$^{a36}$   Also at Max Planck Institute, Munich, Germany, Alexander von Humboldt Research Award\\
% ...
$^{b6}$  Supported by the Deutsche Forschungsgemeinschaft \\
% ...
$^{b16}$   Supported by the Israel Science Foundation\\
$^{b17}$   Supported by the Italian National Institute for Nuclear Physics (INFN) \\
% ...
$^{b27}$   Supported by the Polish Ministry of Science and Higher Education as a scientific project (2009-2010)\\
% ...

}
}

\end{flushleft}

\newpage

\section{Introduction}
\label{sec:intro}

In the Standard Model (SM) events containing an isolated
electron\footnote{Here and in the following, the term ``electron''
denotes generically both the electron and the positron.} or muon of
high transverse momentum, $P_{T}$, in coincidence with large missing
transverse momentum, $P_{T}^{\rm miss}$, arise from the production of
single $W$ bosons with subsequent decay to leptons.
Events of this topology have been observed at the electron--proton
collider HERA~\cite{h1isol1998,zeusisol2000,h1isol2003,zeustop2003}.
An excess of events containing in addition a hadronic final state of
high transverse momentum, $P_{T}^{X}$, was previously reported by the
H1 collaboration in $105$~pb$^{-1}$ of $e^{+}p$
data~\cite{h1isol2003}.
Both the H1 and ZEUS collaborations have recently performed a search
for such events using their complete $e^{\pm}p$ high energy data,
corresponding to an integrated luminosity of approximately
$0.5$~fb$^{-1}$ per experiment~\cite{zeusisol09,h1isol09}.
The event yields are found to be in good overall agreement with the SM
and a measurement of single $W$ production is performed by both
collaborations.
An excess of events is however still seen by H1 at high
$P_{T}^{X}>25$~GeV in the $e^{+}p$ data sample, where $17$ events are
observed compared to a SM prediction of $8.0\pm1.3$~\cite{h1isol09}.


This paper presents a combined analysis of the H1 and ZEUS data,
performed in a common phase space.
The analysis makes use of the full data samples available to both
experiments allowing a more accurate measurement, as well as a more
stringent examination of the high $P_{T}^{X}$ region.
Total event yields and kinematic distributions of events containing an
isolated electron or muon of high transverse momentum and missing
transverse momentum are compared to the SM.
In addition, total and differential cross sections for single $W$
production are measured.


The analysed data were collected between $1994$ and $2007$ at HERA
using the H1 and ZEUS detectors.
The electron and proton beam energies were $27.6$~GeV and $820$~GeV or
$920$~GeV respectively, corresponding to centre--of--mass energies,
$\sqrt{s}$, of $301$~GeV or $319$~GeV.
The data correspond to an integrated luminosity of $0.98$~fb$^{-1}$
comprising $0.39$~fb$^{-1}$ of $e^{-}p$ collisions and
$0.59$~fb$^{-1}$ of $e^{+}p$ collisions, with $9\%$ of the total
integrated luminosity collected at $\sqrt{s}=301$~GeV.
Data collected from $2003$ onwards were taken with a longitudinally
polarised lepton beam, with polarisation typically at a level of
$35\%$.
The residual polarisation of the combined left--handed and
right--handed data periods is less than $3\%$ for both experiments.

\section{Standard Model Processes}
\label{sec:sm}

In this analysis, SM processes are considered signal if they produce
events containing a high $P_{T}$ isolated charged lepton and at least
one high $P_{T}$ neutrino, which escapes detection and leads to
$P_{T}^{\rm miss}$ in the final state.
The production of single $W$ bosons with subsequent decay to an
electron or a muon, which includes a contribution from leptonic
tau--decay, is the main signal contribution to the SM expectation.
The EPVEC~\cite{epvec} Monte Carlo (MC) event generator is used to
calculate the single $W$ production cross section.
The $ep \rightarrow eWX$ events from EPVEC are weighted by a factor
dependent on the transverse momentum and rapidity of the $W$, such
that the resulting cross section corresponds to a calculation
including Quantum Chromodynamics (QCD) corrections at
next--to--leading order (NLO)~\cite{nloepvec}.
The estimated uncertainty on this calculation is $15\%$, which arises
from the uncertainties in the parton densities and the scale
at which the calculation is performed.
The contribution of $ep \rightarrow \nu_eWX$ events to the total
single $W$ production cross section is approximately $7\%$.
The process $ep \rightarrow eZ(\rightarrow \nu\bar{\nu})X$ also
produces high $P_{T}$ isolated electrons and large $P_{T}^{\rm{miss}}$
in the final state.
The visible cross section for this process as calculated by EPVEC is
less than $3\%$ of the predicted single $W$ production cross section
and is neglected in the ZEUS part of the analysis.


All other SM processes are defined as background and contribute to the
selected sample mainly through misidentification or mismeasurement.
Neutral current (NC) deep inelastic scattering (DIS) events ($ep
\rightarrow eX$), in which genuine isolated high $P_{T}$ electrons are
produced, form a significant background in the electron channel when
fake $P_{T}^{\rm{miss}}$ arises from mismeasurement.
Charged current (CC) DIS events ($ep \rightarrow \nu_{e}X$), in which
there is real $P_{T}^{\rm{miss}}$ due to the escaping neutrino,
contribute to the background when fake isolated electrons or muons are
observed.
Lepton pair production ($ep \rightarrow e \ell^{+}\ell^{-}X$)
contributes to the background via events where one lepton escapes
detection and/or measurement errors cause apparent missing momentum.
A small contribution to the background in the electron channel arises
from QED Compton (QEDC) events ($ep \rightarrow e\gamma X$) when
mismeasurement leads to apparent missing momentum.
The background contribution to the analysis from photoproduction is
negligible.

\section{Experimental Method}
\label{sec:exp}

The H1 and ZEUS detectors are general purpose instruments which
consist of tracking systems surrounded by electromagnetic and hadronic
calorimeters and muon detectors, ensuring close to $4\pi$ coverage of
the $e^{\pm}p$ interaction point.
The origin of the coordinate system is the nominal $e^{\pm}p$
interaction point, with the direction of the proton beam defining the
positive $z$--axis (forward region).
The $x$--$y$ plane is called the transverse plane and $\phi$ is the
azimuthal angle.
The pseudorapidity $\eta$ is defined as $\eta = - \ln \tan(
\theta/2)$, where $\theta$ is the polar angle.
Detailed descriptions of the detectors can be found
elsewhere~\cite{h1det,zeusdet}.


The event selection for isolated electrons or muons and missing
transverse momentum is based on those used by the H1~\cite{h1isol09}
and ZEUS~\cite{zeusisol09} experiments.
For the combined analysis, a common phase space is chosen in a region
where both detectors have a high and well understood acceptance.
The event selection for the electron and muon channels is summarised
in Table~\ref{tab:isolepcutstable}, and uses the variables described
below.


Leptons are identified according to the selection criteria employed by
the individual experiments~\cite{h1isol09,zeusisol09}.
Electron candidates are identified as compact and isolated energy
deposits in the electromagnetic calorimeters associated to a track in
the inner tracking system.
Muon candidates are identified as tracks from the inner tracking
system associated with track segments reconstructed in muon chambers
or energy deposits in the calorimeters compatible with a minimum
ionising particle.
Lepton candidates are required to lie within the polar angle range
$15^{\circ} < \theta_{\ell} < 120^{\circ}$ and to have transverse
momentum, $P_{T}^{\ell}$, greater than $10$~GeV.
The lepton is required to be isolated with respect to jets and other
tracks in the event.
Jets are reconstructed from particles in the event not previously
identified as isolated leptons using an inclusive
$k_{T}$~algorithm~\cite{kt}.
The isolation of the lepton is quantified using the distances in
$\eta$--$\phi$ space to the nearest jet $D(\ell;{\rm jet})>1.0$ and
nearest track $D(\ell;{\rm track})>0.5$.
To ensure that the two channels are exclusive, electron channel events
must contain no isolated muons.


The selected events should contain a large transverse momentum
imbalance $P_{T}^{\rm miss}>12$~GeV.
To ensure a high trigger efficiency, the transverse momentum measured
in the calorimeter, $P_{T}^{\rm calo}$, is also required to be greater
than $12$~GeV.
As muons deposit little energy in the calorimeter, $P_{T}^{\rm calo}$
is similar to $P_{T}^{X}$ in the muon channel and therefore the
$P_{T}^{\rm calo}$ requirement effectively acts as a cut on
$P_{T}^{X}$.
For this reason, the muon channel is restricted to the region
$P_{T}^{X}>12$~GeV.


In order to reduce the remaining SM background, a series of further
cuts are applied as described in Table~\ref{tab:isolepcutstable}.
A measure of the azimuthal balance of the event, $V_{\rm ap}/V_{\rm
p}$, is defined as the ratio of the anti--parallel to parallel
momentum components of all measured calorimetric clusters with respect
to the direction of the total calorimetric transverse
momentum~\cite{Adloff:1999ah}.
The difference in azimuthal angle between the lepton and the direction
of the hadronic system, $\Delta \phi_{\ell-X}$, is used to reject SM
background with back--to--back topologies ($\Delta \phi_{\ell-X} =
180^{\circ}$) like those in NC and lepton pair events.
For events with low hadronic transverse momentum $P_{T}^{X}<1.0$~GeV,
the direction of the hadronic system is not well determined and
$\Delta \phi_{\ell-X}$ is set to zero.
The quantity $\delta_{\rm miss}=2E^{0}_{e}- \sum_i (E^{i} -
P_{z}^{i})$, where the sum runs over all detected particles
and $E^{0}_{e}$ is the electron beam energy,
gives a measure of the longitudinal balance of the event.
For an event where only momentum in the proton direction is
undetected, $\delta_{\rm miss}$ is zero.
Further background rejection in the electron channel is achieved using
${\zeta}^{2}_{e}=4 E_{e}E^{0}_{e} \cos^2 \theta_e/2$, where $E_{e}$ is
the energy of the final state electron.
For NC events, where the scattered electron is identified as the
isolated high transverse momentum electron, ${\zeta}^{2}_{e}$ is equal
to the four momentum transfer squared $Q^{2}_{e}$, as measured by the
electron method~\cite{JacquetBlondel}.
The lepton--neutrino transverse mass, $M_{T}^{\ell\nu}$, calculated
using the vectors of the missing transverse momentum and the isolated
lepton, is used to further reject NC (lepton pair) background in the
electron (muon) channel.


\begin{table}[t]
\renewcommand{\arraystretch}{1.4}
\begin{center}
\begin{tabular}{|c||cc|} \hline
\multicolumn{3}{|c|}{\bf H1+ZEUS Isolated Lepton {\boldmath $+ P_{T}^{\rm miss}$} Event Selection}\\
\hline
\multicolumn{1}{|c||}{{\bf Channel}}
& \multicolumn{1}{c}{Electron}
& \multicolumn{1}{c|}{Muon} \\
\hline
\hline
{\bf Basic Event}
& \multicolumn{2}{c|}{$15^\circ<\theta_\ell<120^\circ$}\\
{\bf Selection}
& \multicolumn{2}{c|}{$P_{T}^{\ell}>10$ GeV}\\
& \multicolumn{2}{c|}{$P_{T}^{\rm miss}>12$ GeV}\\ 
& \multicolumn{2}{c|}{$P_{T}^{\rm calo}>12$ GeV}\\
\hline
{\bf Lepton Isolation}
& \multicolumn{2}{c|}{$D(\ell;{\rm jet})>1.0$}\\
& $D(e;{\rm track})>$ $0.5$ for $\theta_{e}>45^\circ$
& $D(\rm{\mu;track})>0.5$ \\
\hline
{\bf Background}
& \multicolumn{2}{c|}{$V_{\rm ap}/V_{\rm p}<$ $0.5$}\\
{\bf Rejection}
& $V_{\rm ap}/V_{\rm p}<0.15$ for $P_{T}^{e}<25$ GeV
& $V_{\rm ap}/V_{\rm p}<0.15$ for $P_{T}^{\rm calo} <25$ GeV\\
& $\Delta\phi_{e-X}<160^\circ$
& $\Delta\phi_{\mu-X}<170^\circ$\\
& $5<\delta_{\rm miss}<50$ GeV
& -- \\
& $\zeta^{2}_{e}>5000$ GeV$^{2}$ for $P_{T}^{\rm calo} <25$ GeV
& -- \\
& \multicolumn{2}{c|}{$M_{T}^{\ell\nu}>10$ GeV}\\
& --
& {$P_{T}^{X}>12$ GeV}\\
& \# electrons $<3$
& -- \\
\hline
\end{tabular}
\end{center}
  \caption{Selection requirements for the electron and muon channels
  in the search for events with an isolated lepton and missing
  transverse momentum.}
\label{tab:isolepcutstable}
\end{table}


The lepton polar--angle acceptance, which is the same as that used in
the ZEUS publication~\cite{zeusisol09}, is the main difference in the
event selection with respect to the published H1 analysis, where
isolated leptons are accepted in the range $5^{\circ} < \theta_{\ell}
< 140^{\circ}$~\cite{h1isol09}.
Additionally, the more restrictive cuts on $\delta_{\rm miss}$ and
$V_{\rm ap}/V_{\rm p}$ are taken from the ZEUS
analysis~\cite{zeusisol09}.
The minimum lepton--neutrino transverse mass and electron multiplicity
requirements are taken from the H1 analysis~\cite{h1isol09}.
The overall H1(ZEUS) efficiency in the common phase--space analysis to
select SM $W \rightarrow e \nu$ events is $30\%$ ($31\%$) and to
select SM $W \rightarrow \mu \nu$ events is $11\%$ ($9\%$), calculated
using EPVEC.


The combination of the H1 and ZEUS results is performed by adding both
the data and MC distributions bin by bin.
The theoretical uncertainty of $15\%$ on single $W$ production from
the reweighted EPVEC prediction is treated as correlated between the
experiments and dominates the SM prediction uncertainty.
Dedicated studies of the significant SM background contributions are
performed by both experiments, using background--enriched control
samples.
The systematic uncertainties attributed to the SM background processes
are derived from the level of agreement between the data and the SM
predictions in these control samples.
Experimental systematic uncertainties, as well as the uncertainties on
the SM background, are treated as uncorrelated between the
experiments.
The systematic uncertainties determined in the combined analysis are
found to be the same as those derived by the individual experiments.
A detailed list of the systematic uncertainties considered can be
found in the respective publications~\cite{h1isol09,zeusisol09}.

\section{Results}
\label{sec:results}

The event yields of the combined H1 and ZEUS search for events
containing an isolated lepton and missing transverse momentum are
summarised in Table~\ref{tab:rates}.
Results are shown for the electron and muon channels separately as
well as combined, for the $e^{+}p$ data, $e^{-}p$ data and the full
HERA $e^{\pm}p$ data.
The results are shown for the full selected sample and for a subsample
at $P_{T}^{X}>25$~GeV.


The signal contribution to the SM expectation, dominated by single $W$
production, is $74\%$ in the combined electron and muon channels for
the full HERA $e^{\pm}p$ data.
The H1 and ZEUS parts of the analysis contribute similarly to the
total signal expectation.
The contribution from signal processes to the total H1 (ZEUS) SM
expectation in the electron channel is $76\%$ ($65\%$) and in the muon
channel $93\%$ ($83\%$).


In the $e^{+}p$ data, $37$ electron events and $16$ muon events are
observed compared to SM predictions of $38.6 \pm 4.7$ and
$11.2 \pm 1.6$ respectively.
In the $e^{-}p$ data, $24$ electron events and $4$ muon events are
observed compared to SM predictions of $30.6 \pm 3.6$ and
$7.4 \pm 1.1$ respectively.
Eleven events in the H1 publication~\cite{h1isol09} are not in the
common phase space: nine events (eight in the electron channel and one
in the muon channel) have $\theta_{\ell}<15^{\circ}$ and two
additional electron channel events fail the stricter $\delta_{\rm
miss}$ condition.
With respect to the published ZEUS analysis~\cite{zeusisol09}, one
event is not in the common phase space due to the cut on transverse
mass.
All twelve events rejected in the combined analysis analysis exhibit
$P_{T}^{X}<25$~GeV.


At large hadronic transverse momentum $P_{T}^{X} >25$~GeV, a total of
$29$ events are observed in the complete HERA $e^{\pm}p$ data compared
to a SM prediction of $24.0 \pm 3.2$.
In the $e^{+}p$ data alone, where an excess of data over the SM
is reported in the H1 analysis~\cite{h1isol09}, $23$ events are
observed with $P_{T}^{X} >25$~GeV compared to a SM prediction of
$14.0 \pm 1.9$.
Seventeen of these $23$ data events are observed in the H1 data,
compared to a SM expectation of $6.7 \pm 1.1$.

\begin{figure}[h]
  \begin{center}
  \vspace{-0.2cm}
      \bf{\textsf{\Large Events with an Isolated Lepton and \boldmath{$\sf P_{\sf T}^{\sf miss}$} at HERA}}\\
      \includegraphics[width=0.49\textwidth]{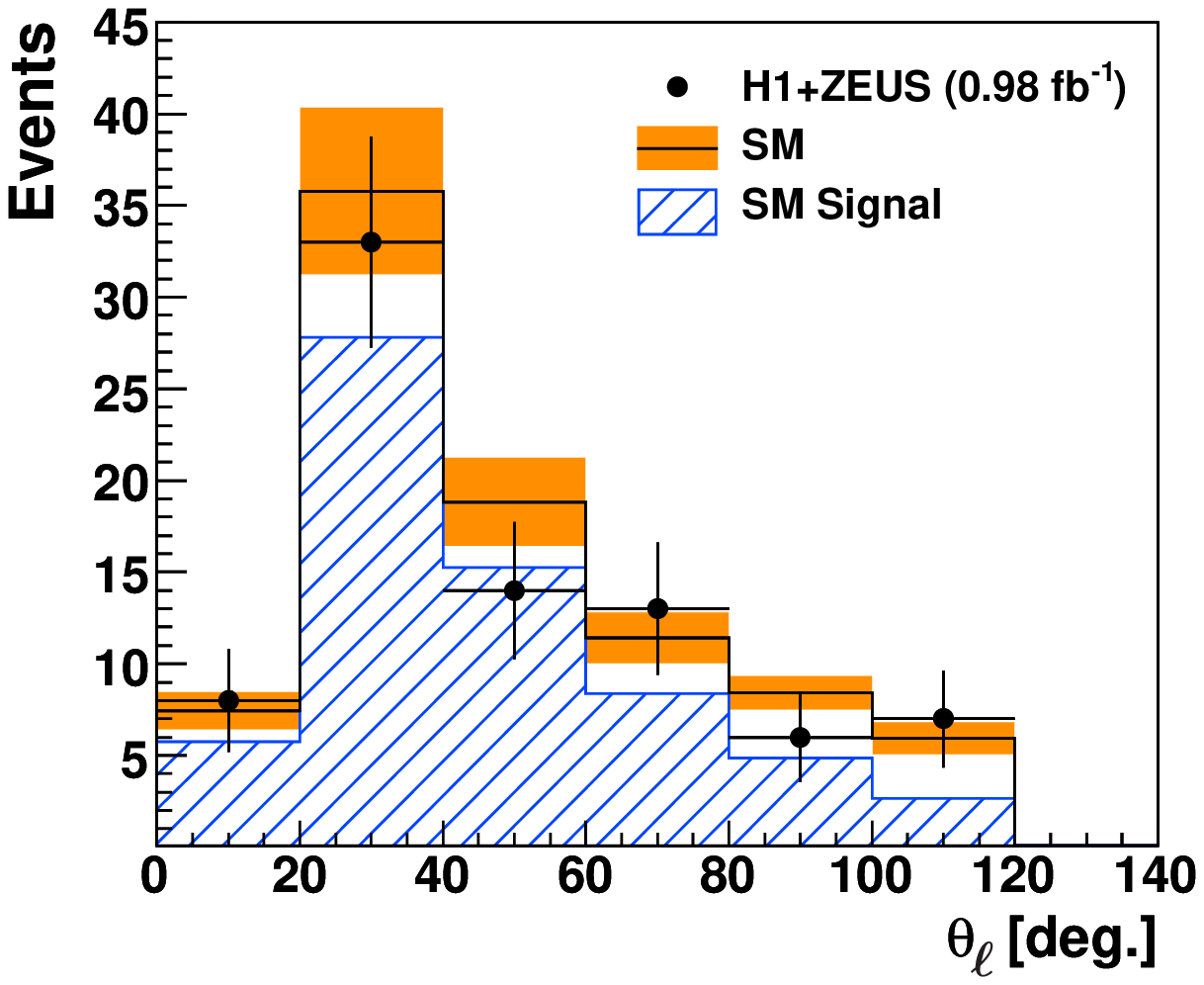}
      \includegraphics[width=0.49\textwidth]{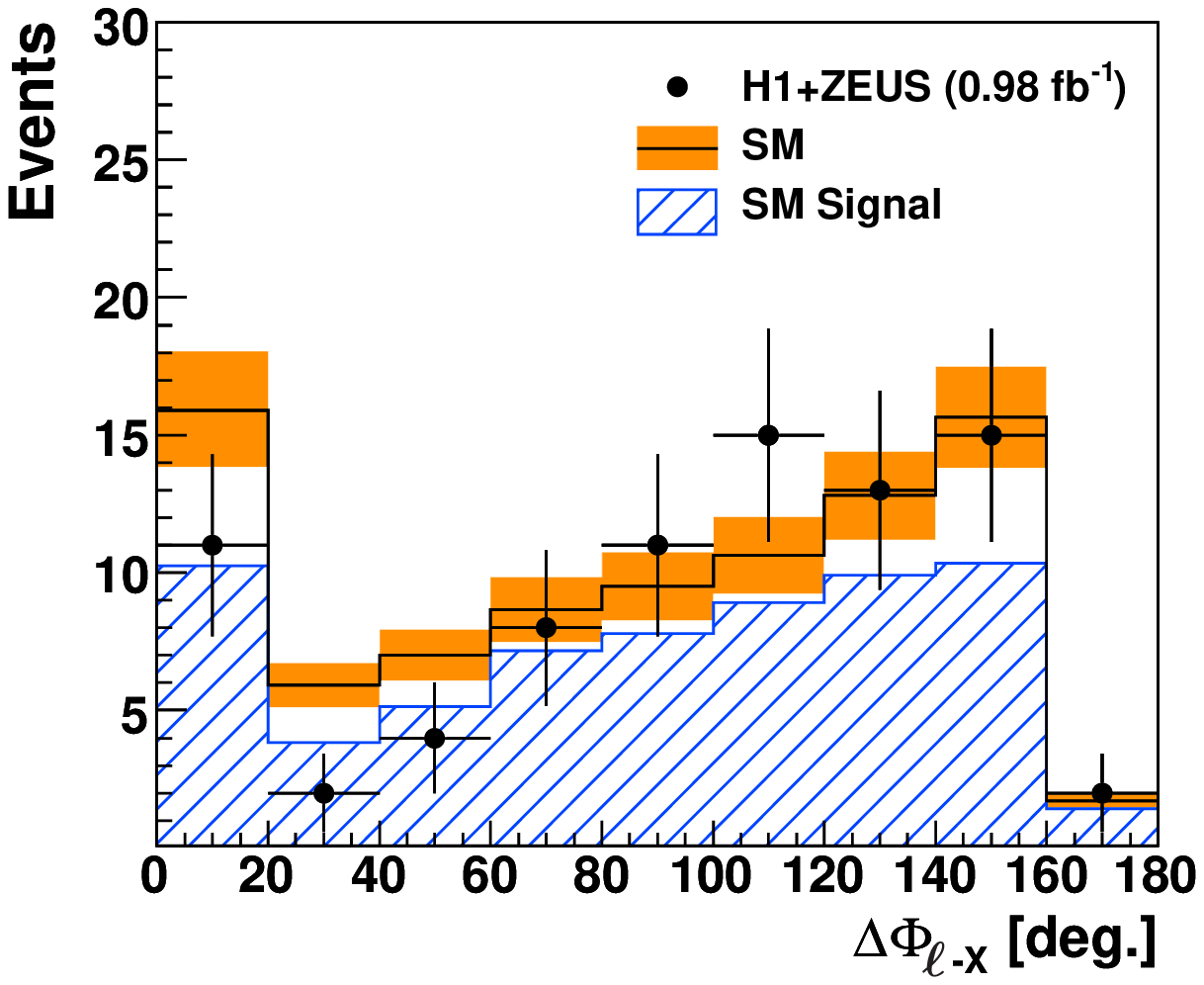}\\
  \vspace{-0.2cm}
      \includegraphics[width=0.49\textwidth]{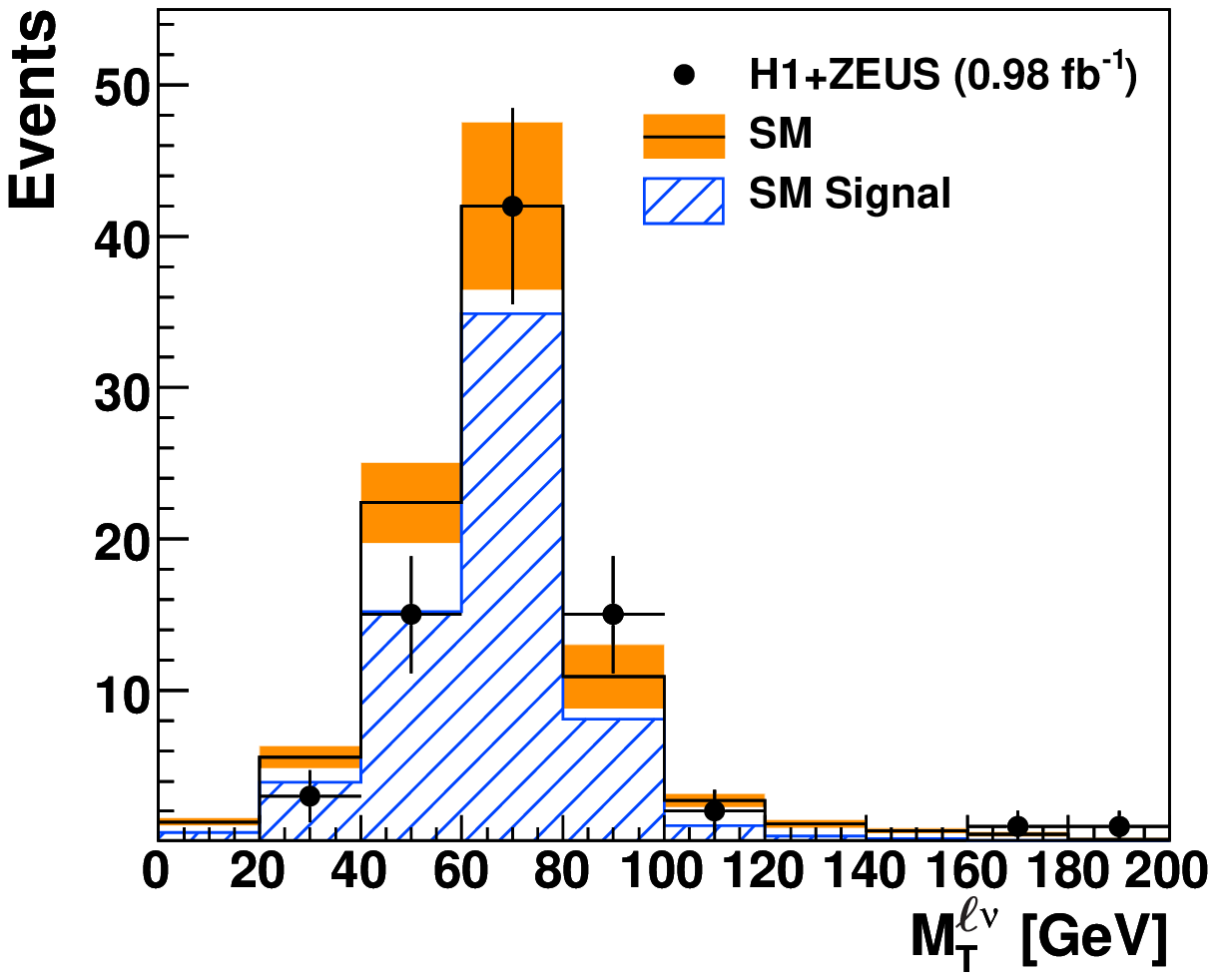}
      \includegraphics[width=0.49\textwidth]{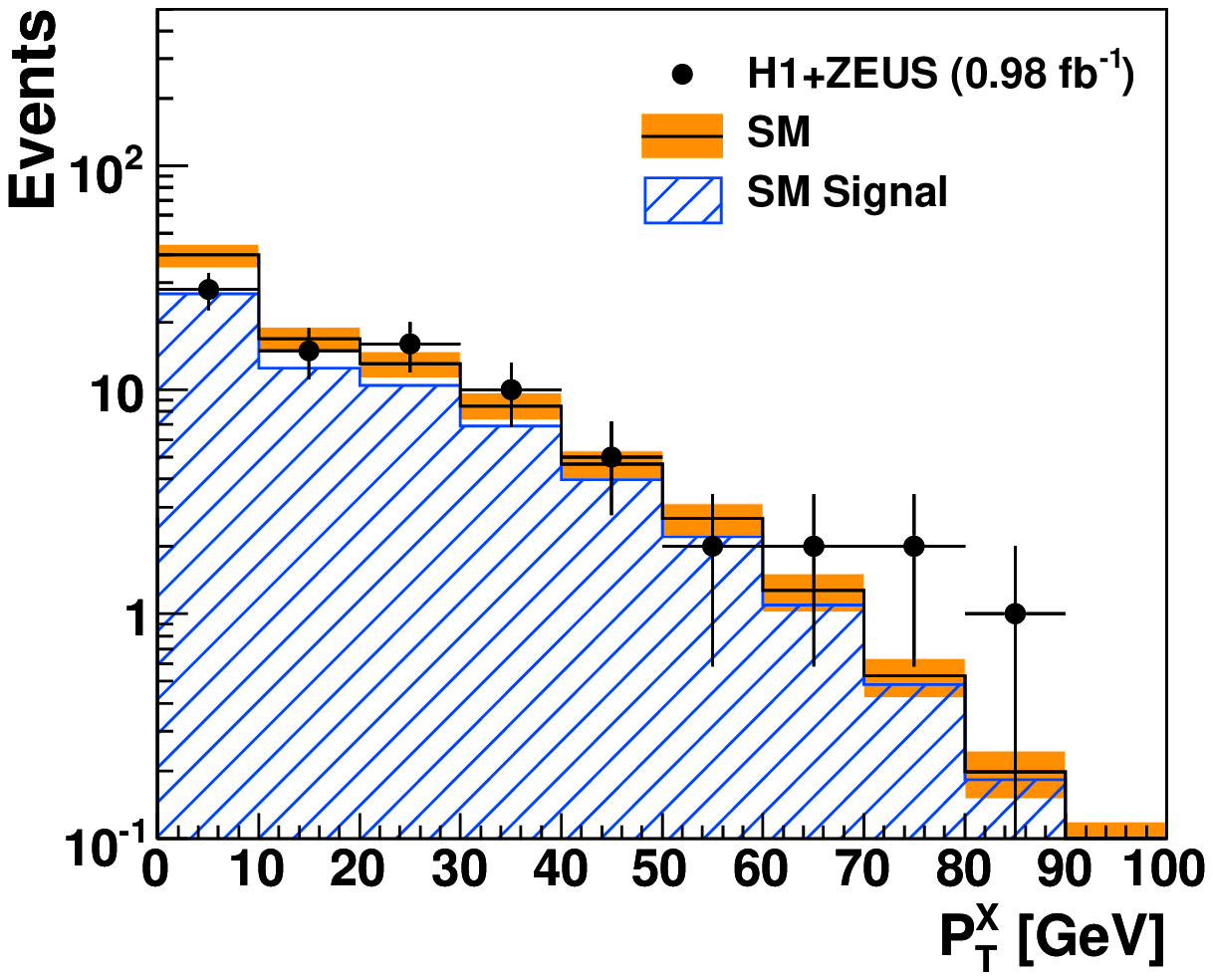}\\
  \vspace{-0.2cm}
      \includegraphics[width=0.49\textwidth]{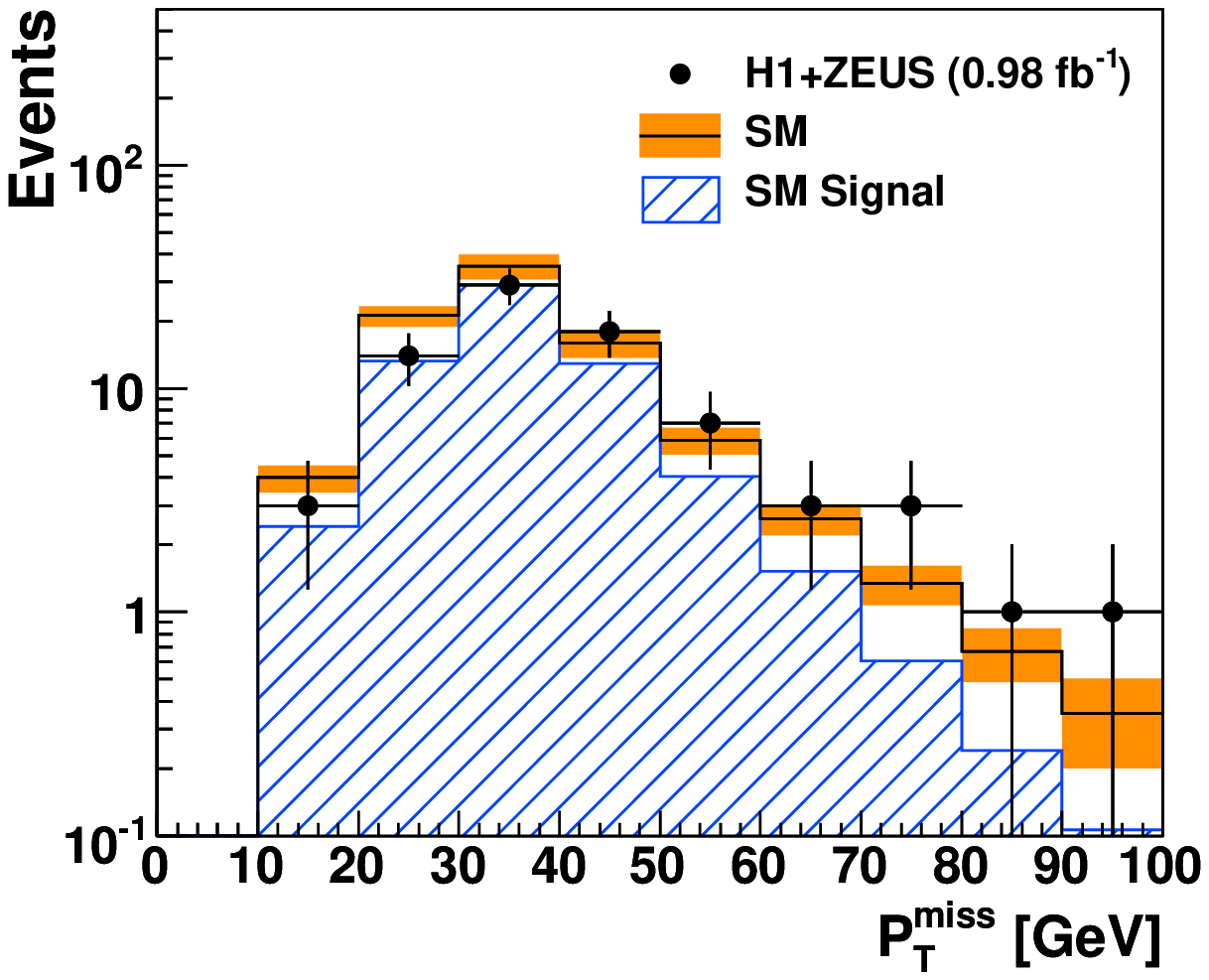}
      \includegraphics[width=0.49\textwidth]{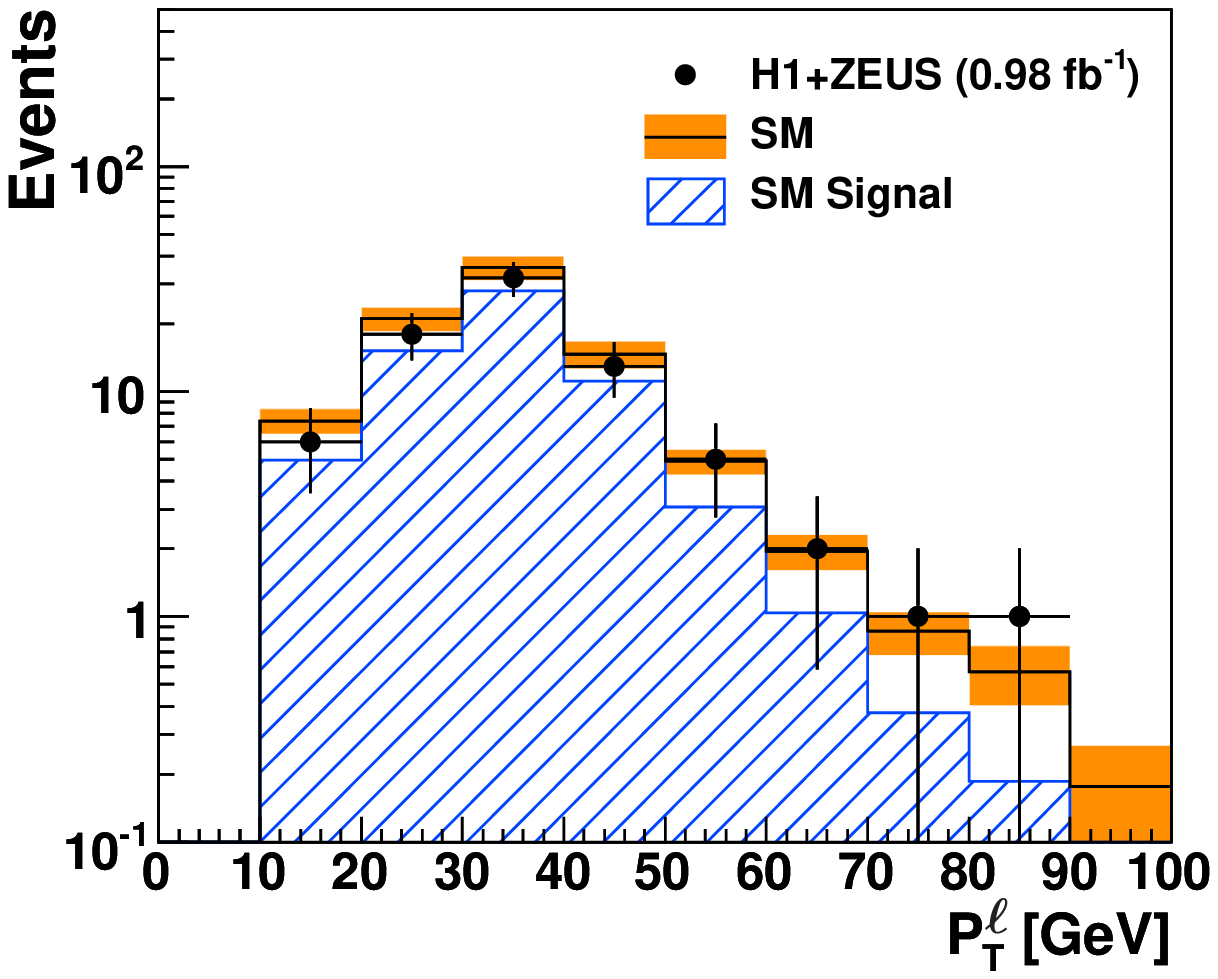}
  \end{center}
  \begin{picture} (0.,0.) 
    \setlength{\unitlength}{1.0cm}
    \put (1.55,18.8){(a)} 
    \put (9.55,18.8){(b)} 
    \put (1.55,12.5){(c)} 
    \put (9.55,12.5){(d)} 
    \put (1.55,6.2){(e)} 
    \put (9.55,6.2){(f)} 
  \end{picture} 
  \vspace{-1.3cm}

  \caption{Distributions of kinematic variables of events with an
  isolated electron or muon and missing transverse momentum in the
  full HERA $e^{\pm}p$ data. Shown are: the polar angle of the lepton
  $\theta_{\ell}$~(a), the difference in the azimuthal angle of the
  lepton and the hadronic systems $\Delta\phi_{\ell-X}$~(b), the
  lepton--neutrino transverse mass $M_{T}^{\ell\nu}$~(c), the hadronic
  transverse momentum $P_{T}^{X}$~(d), the missing transverse momentum
  $P_{T}^{\rm miss}$~(e) and the transverse momentum of the lepton
  $P_{T}^{\ell}$~(f). The data (points) are compared to the SM
  expectation (open histogram). The signal component of the SM
  expectation, dominated by single $W$ production, is shown as the
  hatched histogram. The total uncertainty on the SM expectation is
  shown as the shaded band.}

  \label{fig:isolepfinalsample}
\end{figure}
\begin{figure}[h]
  \begin{center}
      \bf{\textsf{\Large Events with an Isolated Lepton and \boldmath{$\sf P_{\sf T}^{\sf miss}$} at HERA}}\\
      \includegraphics[width=0.495\textwidth]{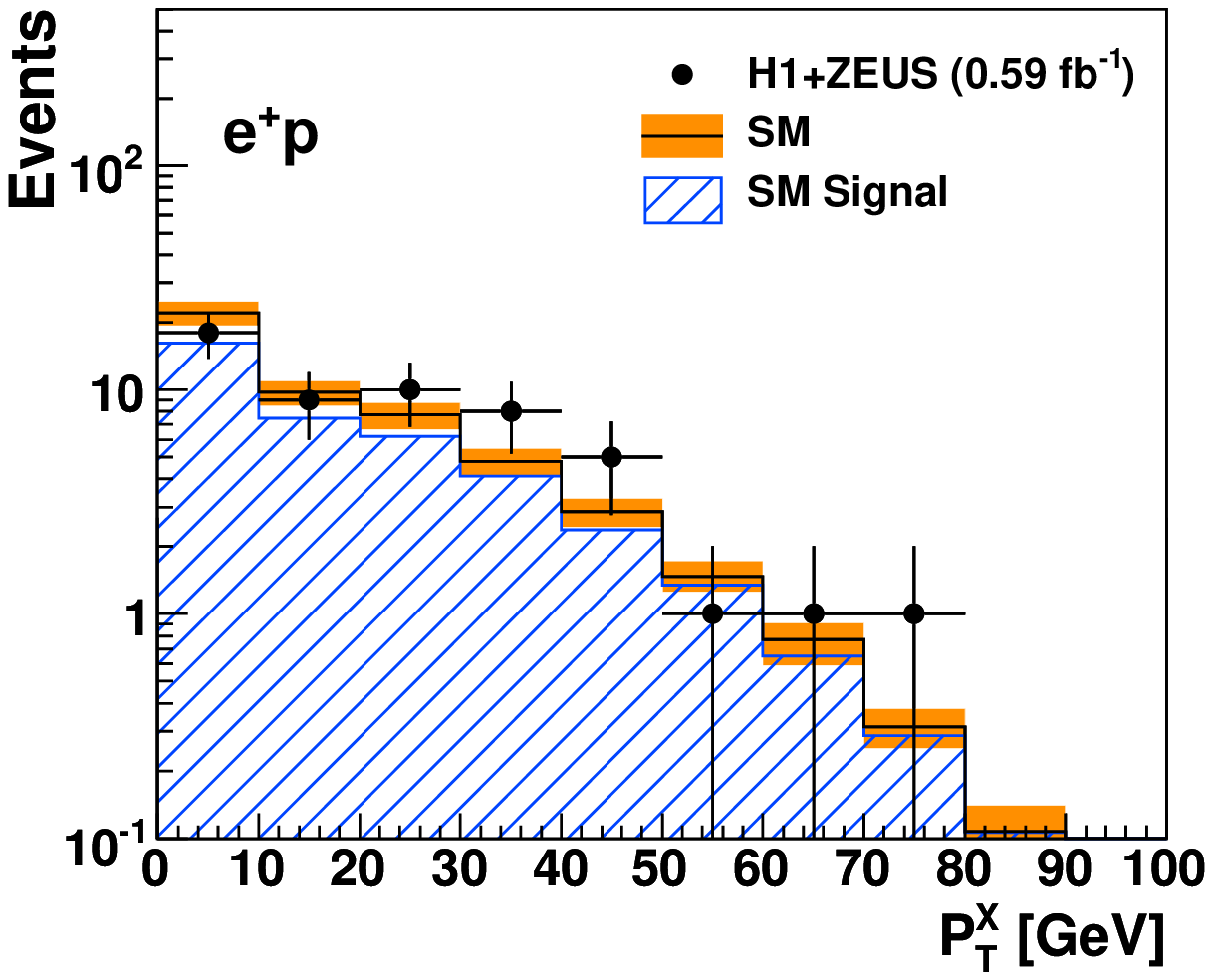}
      \includegraphics[width=0.495\textwidth]{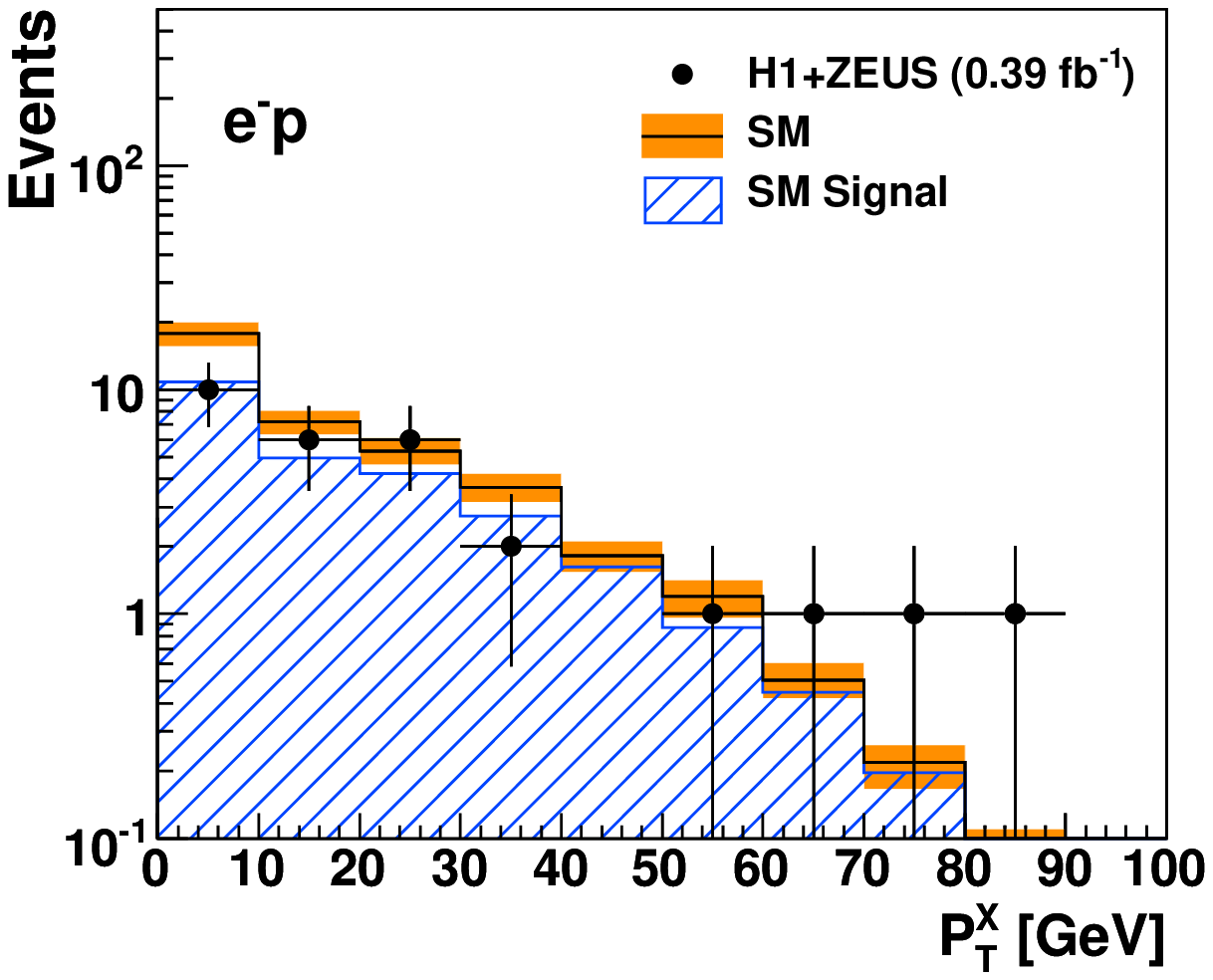}
  \end{center}
  \begin{picture} (0.,0.)
    \setlength{\unitlength}{1.0cm}
    \put ( 6.3,4.8){(a)} 
    \put (14.2,4.8){(b)} 
  \end{picture}  
  \vspace{-1cm}
  \caption{Distributions of the hadronic transverse momentum
  $P_{T}^{X}$ of events with an isolated electron or muon and missing
  transverse momentum for the $e^{+}p$~(a) and $e^{-}p$~(b) HERA data.
  The data (points) are compared to the SM expectation (open
  histogram). The signal component of the SM expectation, dominated by
  single $W$ production, is shown as the hatched histogram. The total
  uncertainty on the SM expectation is shown as the shaded band.}
 \label{fig:isolepptxs}
\end{figure}

Fig.~\ref{fig:isolepfinalsample} shows kinematic distributions of
the complete HERA $e^{\pm}p$ data for the combined electron and muon
channels.
The data are in good agreement with the SM prediction, dominated by
single $W$ production.
The distribution of the lepton polar angle, $\theta_{\ell}$, shows
that the identified lepton is produced mainly in the forward
direction.
The first bin of the $\Delta\phi_{\ell-X}$ distribution is mainly
populated by events with very low values of $P_{T}^{X}$.
The shape of the transverse mass $M_{T}^{\ell\nu}$ distribution shows
a Jacobian peak as expected from single $W$ production.
The observed $P_{T}^{X}$, $P_{T}^{\rm miss}$ and $P_{T}^{\ell}$
distributions are also indicative of single $W$ production, where the
decay products of the $W$ peak around $40$~GeV and the hadronic final
state has typically low $P_{T}^{X}$.
Fig.~\ref{fig:isolepptxs} shows the $P_{T}^{X}$ distribution
separately for the combined $e^{+}p$ and $e^{-}p$ data.


The total and differential single $W$ production cross sections are
evaluated bin by bin from the number of observed events, subtracting
the number of background events, and taking into account the
acceptance and luminosity of the two experiments.
The acceptance, defined as the number of $W$ events reconstructed in a
bin divided by the number of events generated in that bin, is
evaluated using EPVEC and is used to extrapolate the measured cross
section to the full phase space.
The acceptances for the two experiments are found to be similar in
each $P_{T}^{X}$ bin and vary between $27\%$ and $37\%$ in the
electron channel and between $18\%$ and $38\%$ in the muon channel.
The purity of the cross section measurement is greater than $70\%$ in
all bins and is also found to behave similarly for the two
experiments.
For $P_{T}^{X}<12$~GeV, the electron measurement is used to estimate
the muon cross section under the assumption of lepton universality.
Leptonic tau decays from $W \rightarrow \tau \nu$ events are taken into
account in the cross section calculation.
The cross sections are quoted at the luminosity--weighted mean
centre--of--mass energy $\sqrt{s}=317$~GeV of the complete HERA data.

\begin{figure}[h]
  \begin{center}
      \includegraphics[width=0.7\textwidth]{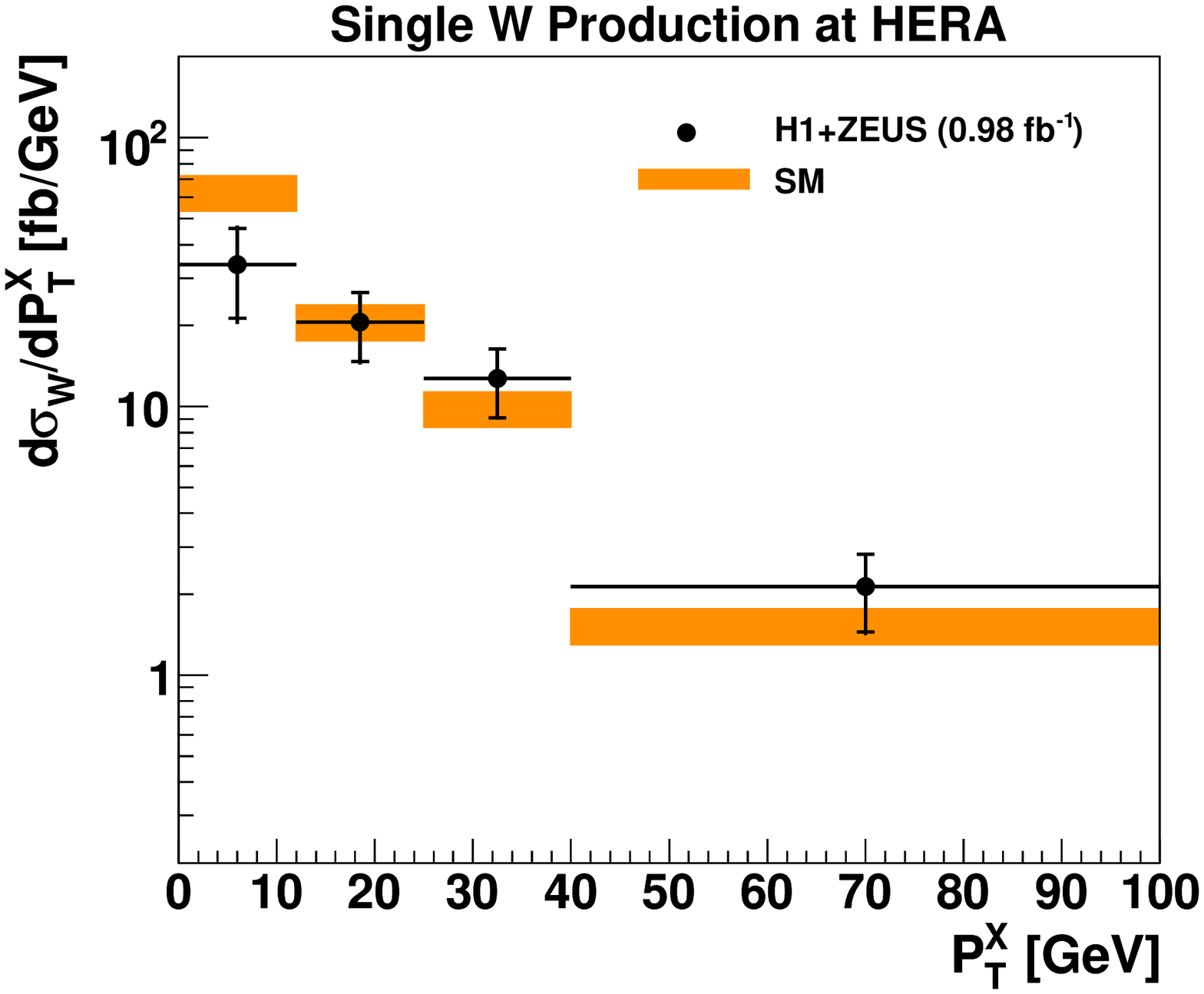}
  \end{center}
  \vspace{-1.0cm}
  \caption{The single $W$ production cross section as a function of
  the hadronic transverse momentum, $P_{T}^{X}$, measured using the
  combined H1 and ZEUS data at a centre--of--mass energy of
  $\sqrt{s}=317$~GeV. The inner error bar represents the statistical
  error and the outer error bar indicates the statistical and
  systematic uncertainties added in quadrature. The shaded band
  represents the uncertainty on the SM prediction.}
 \label{fig:wxs}
\end{figure}

The total single $W$ boson production cross section at HERA is
measured as:
\[
1.06 \pm 0.16~{\rm (stat.)} \pm 0.07~{\rm (sys.)}~{\rm pb},
\]
which agrees well with the SM prediction of $1.26 \pm 0.19$~pb.
The measured differential cross sections, in bins of $P_{T}^{X}$,
are shown in Fig.~\ref{fig:wxs} and given in Table~\ref{tab:xs}.
The differential cross section agrees well with the SM prediction.

\section{Conclusions}
\label{sec:con}

A search for events containing an isolated electron or muon and large
missing transverse momentum produced in $e^{\pm}p$ collisions is
performed with the H1 and ZEUS detectors at HERA in a common phase
space.
The full HERA $e^{\pm}p$ high energy data sample from both experiments
is analysed, corresponding to a total integrated luminosity of
$0.98$~fb$^{-1}$.
A total of $81$ events are observed in the data, compared to a SM
prediction of $87.8 \pm 11.0$.
In the $e^{+}p$ data, at large hadronic transverse momentum $P_{T}^{X}
>25$~GeV, a total of $23$ data events are observed compared to a SM
prediction of $14.0 \pm 1.9$.
The total and differential single $W$ production cross sections are
measured and are found to be in agreement with the SM predictions.

\section*{Acknowledgements}

We are grateful to the HERA machine group whose outstanding efforts
have made these experiments possible. We appreciate the contributions
to the construction and maintenance of the H1 and ZEUS detectors of
many people who are not listed as authors. We thank our funding
agencies for financial support, the DESY technical staff for
continuous assistance and the DESY directorate for their support and
for the hospitality they extended to the non-DESY members of the
collaborations.




\clearpage

\begin{table}[]
\begin{center}
 \renewcommand{\arraystretch}{1.45} 
 \begin{tabular}{|cc||c|rcl||rcl|rcl|}
  \hline
   \multicolumn{2}{|l||}{\bf H1+ZEUS} & Data & \multicolumn{3}{|c||}{SM} & \multicolumn{3}{c|}{SM} & \multicolumn{3}{|c|}{Other SM} \\
   \multicolumn{2}{|l||}{1994--2007 $e^{+}p$ ~~~ $0.59$~fb$^{-1}$} &      & \multicolumn{3}{|c||}{Expectation} & \multicolumn{3}{c|}{Signal} & \multicolumn{3}{|c|}{Processes} \\
   \hline
   \hline
   Electron & Total   & $37$ & $38.6$ & $\pm$ & ~~$4.7$ & $28.9$ & $\pm$ & $4.4$ &  $9.7$ & $\pm$ & $1.4$ \\
   \cline{2-12}
   & $P^X_T > 25$ GeV & $12$ &  $7.4$ & $\pm$ & ~~$1.0$ &  $6.0$ & $\pm$ & $0.9$ &  $1.5$ & $\pm$ & $0.3$ \\
   \hline
   Muon     & Total   & $16$ & $11.2$ & $\pm$ & ~~$1.6$ &  $9.9$ & $\pm$ & $1.6$ &  $1.3$ & $\pm$ & $0.3$ \\
   \cline{2-12}
   & $P^X_T > 25$ GeV & $11$ &  $6.6$ & $\pm$ & ~~$1.0$ &  $5.9$ & $\pm$ & $0.9$ &  $0.8$ & $\pm$ & $0.2$ \\
   \hline
   Combined & Total   & $53$ & $49.8$ & $\pm$ & ~~$6.2$ & $38.8$ & $\pm$ & $5.9$ & $11.1$ & $\pm$ & $1.5$ \\
   \cline{2-12}
   & $P^X_T > 25$ GeV & $23$ & $14.0$ & $\pm$ & ~~$1.9$ & $11.8$ & $\pm$ & $1.9$ &  $2.2$ & $\pm$ & $0.4$ \\
   \hline

   \multicolumn{12}{c}{}\\

   \hline
   \multicolumn{2}{|l||}{\bf H1+ZEUS} & Data & \multicolumn{3}{|c||}{SM} & \multicolumn{3}{c|}{SM} & \multicolumn{3}{|c|}{Other SM} \\
   \multicolumn{2}{|l||}{1998--2006 $e^{-}p$ ~~~ $0.39$~fb$^{-1}$} &      & \multicolumn{3}{|c||}{Expectation} & \multicolumn{3}{c|}{Signal} & \multicolumn{3}{|c|}{Processes} \\
   \hline
   \hline
   Electron & Total   & $24$ & $30.6$ & $\pm$ & ~~$3.6$ & $19.4$ & $\pm$ & $3.0$ & $11.2$ & $\pm$ & $1.9$ \\
   \cline{2-12}
   & $P^X_T > 25$ GeV &  $4$ &  $5.6$ & $\pm$ & ~~$0.8$ &  $4.0$ & $\pm$ & $0.6$ &  $1.6$ & $\pm$ & $0.4$ \\
   \hline
   Muon     & Total   &  $4$ &  $7.4$ & $\pm$ & ~~$1.1$ &  $6.6$ & $\pm$ & $1.0$ &  $0.9$ & $\pm$ & $0.3$ \\
   \cline{2-12}
   & $P^X_T > 25$ GeV &  $2$ &  $4.3$ & $\pm$ & ~~$0.7$ &  $3.9$ & $\pm$ & $0.6$ &  $0.4$ & $\pm$ & $0.2$ \\
   \hline
   Combined & Total   & $28$ & $38.0$ & $\pm$ & ~~$3.4$ & $26.0$ & $\pm$ & $3.4$ & $12.0$ & $\pm$ & $2.0$ \\
   \cline{2-12}
   & $P^X_T > 25$ GeV &  $6$ & $10.0$ & $\pm$ & ~~$1.3$ &  $7.9$ & $\pm$ & $1.2$ &  $2.1$ & $\pm$ & $0.5$ \\
   \hline

   \multicolumn{12}{c}{}\\

   \hline
   \multicolumn{2}{|l||}{\bf H1+ZEUS} & Data & \multicolumn{3}{|c||}{SM} & \multicolumn{3}{c|}{SM} & \multicolumn{3}{|c|}{Other SM} \\
   \multicolumn{2}{|l||}{1994--2007 $e^{\pm}p$ ~~~ $0.98$~fb$^{-1}$} &   & \multicolumn{3}{|c||}{Expectation} & \multicolumn{3}{c|}{Signal} & \multicolumn{3}{|c|}{Processes} \\
   \hline
   \hline
   Electron & Total   & $61$ & $69.2$ & $\pm$ &  ~~$8.2$ & $48.3$ & $\pm$ & $7.4$ & $20.9$ & $\pm$ & $3.2$ \\
   \cline{2-12}
   & $P^X_T > 25$ GeV & $16$ & $13.0$ & $\pm$ &  ~~$1.7$ & $10.0$ & $\pm$ & $1.6$ &  $3.1$ & $\pm$ & $0.7$ \\
   \hline
   Muon     & Total   & $20$ & $18.6$ & $\pm$ &  ~~$2.7$ & $16.4$ & $\pm$ & $2.6$ &  $2.2$ & $\pm$ & $0.5$ \\
   \cline{2-12}
   & $P^X_T > 25$ GeV & $13$ & $11.0$ & $\pm$ &  ~~$1.6$ &  $9.8$ & $\pm$ & $1.6$ &  $1.2$ & $\pm$ & $0.3$ \\
   \hline
   Combined & Total   & $81$ & $87.8$ & $\pm$ & $11.0$  & $64.7$ & $\pm$ & $9.9$ & $23.1$ & $\pm$ & $3.3$ \\
   \cline{2-12}
   & $P^X_T > 25$ GeV & $29$ & $24.0$ & $\pm$ &  ~~$3.2$  & $19.7$ & $\pm$ & $3.1$ &  $4.3$ & $\pm$ & $0.8$ \\
  \hline
 \end{tabular}
\end{center}
\vspace{-0.5cm}
  \caption{Summary of the combined H1 and ZEUS search for events with
  an isolated electron or muon and missing transverse momentum for the
  $e^{+}p$ data (top), $e^{-}p$ data (middle) and the full HERA data
  set (bottom). The results are shown for the full selected sample and
  for the subsample with hadronic transverse momentum
  $P_{T}^{X}>25$~GeV. The number of observed events is compared to the
  SM prediction. The SM signal (dominated by single $W$ production)
  and the total background contribution are also shown. The quoted
  uncertainties contain statistical and systematic uncertainties
  added in quadrature.}
 \label{tab:rates}
\end{table}

\newpage

\begin{table}[]
\renewcommand{\arraystretch}{1.5}
\begin{center}
\begin{tabular}{| r @{$\,-\,$} l | r @{$\,\pm\,$} r @{$\,\pm\,$} l | c | }
  \hline
 \multicolumn{6}{|c|}{ \textbf{H1+ZEUS Differential Single {\boldmath $W$} Production Cross Section} } \\
\hline 
 \multicolumn{2}{|c|}{$P_{T}^{X}$~[GeV] } & Measured & stat. & sys. [fb / GeV] & SM NLO [fb / GeV]\\
\hline
\hline
   $0$ & ~~$12$  &  $33.6$ &   $12.3$ &   $5.0$ &  $62.7$ $\pm$   $9.4$ \\
  $12$ & ~~$25$  &  $20.6$ &    $6.0$ &   $1.9$ &  $20.7$ $\pm$   $3.1$ \\
  $25$ & ~~$40$  &  $12.7$ &    $3.6$ &   $1.0$ &  $~~9.8$ $\pm$  $1.5$ \\
  $40$ &  $100$  &   $2.1$ &    $0.7$ &   $0.2$ &  $~~1.5$ $\pm$  $0.2$ \\
\hline
\end{tabular}
\end{center}
 \caption{The differential single $W$ boson production cross section
 d$\sigma_{W}$/d$P_{T}^{X}$, with statistical (stat.) and systematic
 (sys.) errors, measured using the combined H1 and ZEUS data. The
 cross sections are quoted at a centre--of--mass energy
 $\sqrt{s}=317$~GeV. Also shown are the expectations, including the
 theoretical uncertainties, for the Standard Model calculated at
 next--to--leading order (SM NLO).}
 \label{tab:xs}
\end{table}

\newpage

\newpage

\begin{thebibliography}{99}


\bibitem{h1isol1998}
C.~Adloff {\it et al.} [H1 Collaboration],
Eur.\ Phys.\ J.\ C {\bf 5} (1998) 575 [hep-ex/9806009].


\bibitem{zeusisol2000}
J.~Breitweg {\it et al.} [ZEUS Collaboration],
Phys.\ Lett.\ B {\bf 471} (2000) 411 [hep-ex/9907023].


\bibitem{h1isol2003}
V.~Andreev {\it et al.} [H1 Collaboration],
Phys.\ Lett.\ B {\bf 561} (2003) 241 [hep-ex/0301030].


\bibitem{zeustop2003}
S.~Chekanov {\it et al.} [ZEUS Collaboration],
Phys.\ Lett.\ B {\bf 559} (2003) 153 [hep-ex/0302010].


\bibitem{zeusisol09}
S.~Chekanov {\it et al.} [ZEUS Collaboration],
Phys.\ Lett.\ B {\bf 672} (2009) 106 [arXiv:0807.0589].


\bibitem{h1isol09}
F.~D.~Aaron {\it et al.} [H1 Collaboration],
accepted by Eur.\ Phys.\ J.\ C. [arXiv:0901.0488].


\bibitem{h1det}
I.~Abt {\it et al.} [H1 Collaboration],
Nucl.\ Inst.\ Meth.\ A {\bf 386} (1997) 310;\\
I.~Abt {\it et al.} [H1 Collaboration],
Nucl.\ Inst.\ Meth.\ A {\bf 386} (1997) 348;\\
R.~D.~Appuhn {\it et al.}  [H1 SPACAL Group],
Nucl.\ Instrum.\ Meth.\ A {\bf 386} (1997) 397.


\bibitem{zeusdet}
ZEUS\ Collaboration\ (U.~Holm ed.),\ {\it The ZEUS Detector}.\ Status Report\ (unpublished)\ DESY\ (1993),\ available\ at\ http://www-zeus.desy.de/bluebook/bluebook.html. 


\bibitem{epvec}
U.~Baur, J.~A.~Vermaseren and D.~Zeppenfeld,
Nucl.\ Phys. B {\bf 375} (1992) 3.


\bibitem{nloepvec}
K.~P.~Diener, C. Schwanenberger and M. Spira, Eur. Phys. J. C {\bf 25} (2002)
405 [hep-ph/0203269];\\
P.~Nason,~R.~R\"{u}ckl and M.~Spira, J. Phys. G {\bf 25}, (1999) 1434 [hep-ph/9902296];\\
M. Spira, [hep-ph/9905469].


\bibitem{kt}
S.~D.~Ellis and D.~E.~Soper,
Phys.\ Rev.\ D {\bf 48} (1993) 3160
[hep-ph/9305266];\\
S.~Catani {\it et al.},
Nucl.\ Phys.\ B {\bf 406} (1993) 187.


\bibitem{Adloff:1999ah}
C.~Adloff {\it et al.}  [H1 Collaboration],
Eur.\ Phys.\ J.\ C {\bf 13} (2000) 609
[hep-ex/9908059].


\bibitem{JacquetBlondel}
F.~Jacquet and A.~Blondel,
proceedings of ``Study of an $ep$ Facility for Europe'',
(U.~Amaldi ed.), DESY (1979),
DESY 79/48, 391.

\end{thebibliography}
\end{document}